# Coherently excited Hong-Ou-Mandel effects using frequency-path correlation

Byoung S. Ham

School of Electrical Engineering and Computer Science, Gwangju Institute of Science and Technology
123 Chumdangwagi-ro, Buk-gu, Gwangju 61005, South Korea
(Submitted on July 03, 2023; bham@gist.ac.kr)

**Abstract**
Nonlocal quantum correlation has been the main issue of quantum mechanics over the last century. The Hong-Ou-Mandel (HOM) effect relates to the two-photon intensity correlation on a beam splitter, resulting in a nonclassical photon-bunching phenomenon. The HOM effect has been used to verify the quantum feature via Bell measurements for quantum technologies such as quantum repeaters and photonics quantum computing. Here, a coherence version of the HOM effect is proposed and analyzed to understand the fundamental physics of the anticorrelation and entanglement. For this, frequency-correlated coherent photon pairs are prepared in an independent set of Mach-Zhender interferometers (MZI) using a synchronized pair of modulators from an attenuated laser. For the HOM effect, the phase relation between frequency-correlated photons plays an essential role. For the product-basis randomness, the symmetrically modulated two independent MZIs are combined together incoherently. A classical intensity product between two independent photodetectors is also discussed for the same HOM effect in a selective macroscopic measurement scheme.

1. **INTRODUCTION**

Nonclassical two-photon intensity correlation is the key to quantum information science [1,2]. Since 1987 [3], the Hong-Ou-Mandel (HOM) effect has been intensively studied for the anticorrelation of photon bunching between entangled photons [3-12]. The nonclassical feature of the HOM effect requires indistinguishable photon characteristics on a beam splitter (BS). Here, indistinguishability simply means the wave nature of a photon in quantum mechanics. Based on this understanding, the destructive interference of the HOM effect should confine a relative phase between two impinging photons due to the BS matrix representation rooted in coherence optics [13,14]. Although the quantum operator-based HOM analysis in conventional quantum mechanics explains the destructive interference between two product bases [13], such an understanding of the same phased photons does not come along with the BS matrix, where the BS matrix is a direct result of coherence optics [15]. To understand the HOM effect correctly, thus, the paired photons on a BS have been coherently analyzed for the inherent phase shift between paired photons [16,17]. Recent observations of the HOM effect between independent light sources should give such a coherence clue because an etalon pair initializes the phase relation between all incoherent photon pairs [5,9].

Recently, wave nature-based interpretations of the delayed-choice quantum eraser have been experimentally demonstrated for the fundamental physics of quantum measurements, i.e., violation of the cause-effect relation [18]. In ref. [18], it has been verified that the phase coherence of a photon is the bedrock of the quantum eraser. The nonlocal quantum correlation between space-like separated paired photons [19,20] has also been newly interpreted with the wave nature of a photon [21]. Based on these coherence interpretations of the quantum feature, observations of the HOM effect using different colored lights [9] or independent emitters [5] can also be understood coherently without violating quantum mechanics [16,17]. Here, a coherence model of the HOM effect is proposed and analyzed for the fundamental physics of a fixed phase relation between individually paired photons randomly generated from an attenuated laser. For this, an independent pair of Mach-Zehnder interferometers (MZIs) is used for the coherence HOM model via frequency-modulated two-mode coherent photon pairs. For frequency modulation of coherent photons, synchronized acousto-optic modulators (AOMs) are adopted to generate frequency-path correlated photon pairs. Finally, a coherence solution of the coherence HOM model is derived from a statistical mixture of two independent MZIs, resulting in the same HOM effect. Numerical calculations of the solution show the so-called HOM dip with ensemble degraded heterodyne fringes, as usual. For the coherence HOM effect, an inherent relative phase between paired photons plays an essential role without violating the complementarity theory, as already discussed in the EPR paradox [22-24].

## 2. A SYMMETRIC COHERENCE HOM MODEL

Figure 1 shows the schematic of the proposed coherence HOM model using two independent MZIs via synchronized frequency modulations. Figure 1(a) shows a set of symmetric HOM models for product-basis randomness at the mean value of the second-order intensity correlations in a mixed scheme. Each photon pair is obtained from an attenuated laser by Poisson statistics. The Insets show frequency-modulated photon pair (top) and linear superposition of the product bases from both MZIs (bottom). Figure 1(b) shows the schematic of diffraction angle-independent spatial overlap on a BS. Unlike conventional HOM schemes [1,3,4,6-8] requiring entangled photon pairs generated from a such as spontaneous parametric down-conversion (SPDC) process [25,26], Fig. 1 uses non-quantum particles, as differently demonstrated in refs [5] and [9]. In SPDC-entangled photon pairs, signal and idler photons are symmetrically detuned across the center frequency ($f_0$) according to the phase matching condition of second-order ($\chi^{(2)}$) nonlinear optics (see also the upper Inset) [25-27]. In Fig. 1, each spatially distinguished two-mode coherent photon pair in both MZIs satisfies the symmetric frequency-path correlations. In each MZI, a pair of AOMs is synchronously modulated, resulting in an opposite frequency detuning relation (see Fig. 1(b)).

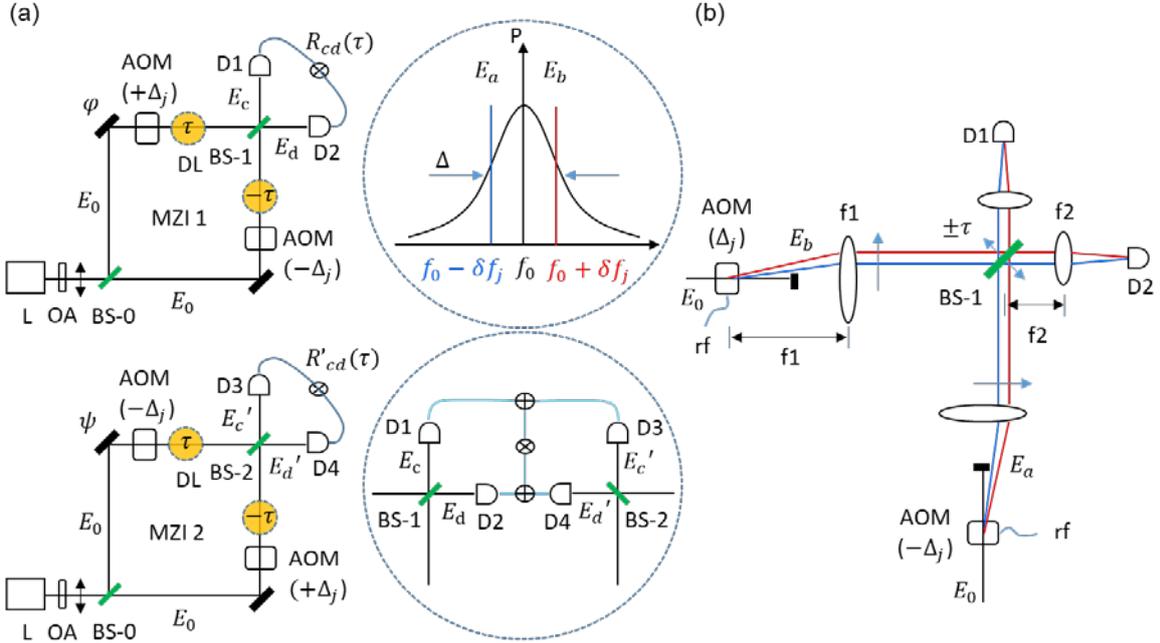

**Fig. 1.** Schematic of a coherent version of the HOM effect based on frequency-modulated coherent photon pairs. (a) Schematic of symmetric coherence HOM. (b) Spatial overlap of the frequency-modulated photon pairs. Inset: (top) frequency modulated photon pairs; (bottom) schematic of coupled MZIs. L: commercial laser, BS: nonpolarizing beam splitter, PBS: polarizing beam splitter, OA: optical attenuator, AOM: acousto-optic modulator, Q: quarter-wave plate, M: mirror, DL: delay line, D: single photon detector, S: beam shutter, η: phase control. Inset: AOM-generated frequency-path correlation pair.

By the coincidence measurements between the output photons from BS-1 (BS-2), only doubly bunched photons from an attenuated laser are post-selected for the HOM effect. Higher-order bunched photons are statistically neglected by Poisson distribution, whose generation rate is ~ 1% [28]. Unpaired single photons or vacuum states are also excluded by the definition of coincidence measurements [28]. By setting the opposite frequency-path correlation between MZI 1 and MZI 2, all post-selected photon pairs should satisfy the same frequency-path correlation relation as the SPDC entangled photons (see the lower Inset). For statistically random measurements, the AOM modulation rate is set to be faster than the bandwidth of the single-photon detectors. The fixed relative phase $\pi/2$ between paired photons is inherently provided by BS-0 [15]. The inherent phase shift $\pm\pi/2$ between entangled photons [16] has already been observed in an entangled ion pair [29]. For symmetric

HOM models in Fig. 1(a), the AOMs are simply swapped in MZI 2, where $\psi = \varphi \pm \pi$ should be kept, too. Here, $\psi = \varphi \pm \pi$ relation is automatically satisfied by the same path-length control, whose resulting phases are opposite to $\pm \delta f_j$. Because most photon pairs are statistically alternative [28], the intensity-product superposition is just measurement addition (linear superposition) from both MZIs.

## 3. ANALYSIS

In the MZI 1 of Fig. 1(a), the output photons from BS-1 can be described using the BS matrix representation [15]:

$$E_c(t) = \frac{E_0}{\sqrt{2}} e^{i(f_0 t - i\Delta_j)} \left(1 - e^{i(\varphi + 2\Delta_j)}\right), \tag{1}$$

$$E_d(\tau) = \frac{iE_0}{\sqrt{2}} e^{i(f_0 t - i\Delta_j)} \left(1 + e^{i(\varphi + 2\Delta_j)}\right), \tag{2}$$

where $\varphi$ is a universal phase control that is independent of the detuning $\pm \delta f_j$ from $f_0$, and $\Delta_j = \delta f_j t$. Here, it should be noted that the sign of $\varphi$ corresponds to the sign of detuning $\delta f_j$. Thus, corresponding mean intensities are as follows:

$$\langle I_c(\tau) \rangle = \frac{I_0}{2N+1} \sum_{j=-N}^{N} [1 - \cos(\varphi + 2\Delta_j)], \tag{3}$$

$$\langle I_d(\tau) \rangle = \frac{I_0}{2N+1} \sum_{j=-N}^{N} [1 + \cos(\varphi + 2\Delta_j)]. \tag{4}$$

In Eqs. (3) and (4), the MZI path-length difference should be within the coherence length. More importantly, $\Delta_j$ is a function of both free running time 't' by the AOM modulators and the delay time $\tau$ by the delay line (DL). The free-running time-based measurement issue can be solved with a gated-measurement technique to freeze 't' (see Methods). For a single-photon counting module, however, it is usually impossible to freeze the free running time due to its averaging method required to reduce a statistical error. Thus, the single photon counting module is set for the gated measurement or simply replaced by a fast digital oscilloscope. On the contrary, SPDC-generated entangled photon pairs have no 't' dependence because of the $\chi^{(2)}$ process, where the ransom phase of the pump photon is acted as a common phase to the sum of entangled photons. By the way, MZI has no difference between a single photon and continuous wave (cw) light, because the interference fringes are rooted in a self-interference of a single photon [30]. In other words, the second-order intensity correlation of the HOM effect is the sum of two coherent photons' self-interferences on a BS. Thus, Fig. 1 should be appropriate for the classical intensity product, too (see below). To have a meaningful change in $\cos(\varphi + 2\Delta_j)$, the order of $\tau$ should be equivalent to $\Delta^{-1}$. As a result, the first-order intensity correlations in MZI 1 show beating fringes as a function of $\tau$. In each MZI, thus, the requirement of uniform intensity in the first-order intensity correlation of the HOM effect is violated.

For the second-order intensity correlation $R_{cd}(\tau)$, the following quantum relation is obtained from Eqs. (1) and (2):

$$\begin{aligned}\sqrt{R_{cd}(\tau)} &= E_c(t+\tau) E_d(t) \\ &= \frac{iI_0}{2} e^{2i(f_0 \tau - i\Delta_j)} \left(1 - e^{i(\varphi + 2\Delta_j)}\right)\left(1 + e^{i(\varphi + 2\Delta_j)}\right) \\ &= \frac{iI_0}{2} e^{2i(f_0 \tau - i\Delta_j)} \left(1 - e^{2i(\varphi + 2\Delta_j)}\right). \end{aligned} \tag{5}$$

Thus, the mean second-order intensity correlation with the gated measurements is as follows:

$$\langle R_{cd}(\tau) \rangle = \frac{I_0^2}{2N+1} \sum_{j=-N}^{N} \sin^2(\varphi + 2\delta f_j \tau). \tag{6}$$

From Eq. (6), the definite condition of the HOM effect for anticorrelation at $\tau = 0$ is $\varphi = 0$. For $\tau \Delta \gg 1$, the average of the sine square term in Eq. (6) becomes saturated into the classical lower bound of $g^{(2)}(\tau) = \frac{\langle R_{cd}(\tau) \rangle}{\langle I_c(t) \rangle \langle I_d(t) \rangle} = 0.5$ [16]. Thus, the inherent phase shift $\pi/2$ between frequency-modulated photon pairs at $\pm \Delta_j$ is confirmed as the origin of the photon bunching phenomenon in the HOM effect. This inherent phase shift for all paired photons in Fig. 1(a) is provided by BS-0 [15]. The fixed relative phase $\pi/2$ has nothing to do with $\delta f_j \tau$. So far, this inherent phase relation between paired photons has never been discussed in the conventional quantum information community, even though it has been observed [29].

Regarding the classical counterpart $\langle P_{cd}(\tau) \rangle = \langle I_c(t+\tau) I_d(t) \rangle$, the same result can be obtained in a pulse scheme by the gated operation: $\langle P_{cd}(\tau) \rangle = \frac{I_0^2}{2N+1} \sum_{j=-N}^{N} sin^2(\varphi + 2\delta f_j \tau)$. For this, single-photon detectors for $I_c$ and $I_d$ can be replaced by conventional PDs. In a direct connection to a fast digital oscilloscope, gated intensity products $\langle P_{cd}(\tau) \rangle$ should show the same feature as $\langle R_{cd}(\tau) \rangle$. Here, the gated measurement is a sort of selective measurement in quantum correlations such as in Franson-type nonlocal correlation [31] and Bell inequality violations [32]. To effectively filter out the free running time 't,' the gated time window must be much shorter than the sinusoidal oscillation in Eqs. (3) and (4). However, this gated time window cannot be shorter than the inverse of the AOM's modulation bandwidth. To remove the coherence effect between consecutive events, a pulse scheme is definitely needed.

In MZI 2 of Fig. 1(a), the AOM's positions are swapped to reverse the frequency-path correlation. For the complete reversing, the $\varphi$ must be reversed, too: $\psi = \varphi \pm \pi$. Thus, the photon fields in MZI 2 are represented as:

$$E_c^{'}(t) = \frac{E_0}{\sqrt{2}} e^{i(f_0\tau + i\Delta_j)} \left(1 - e^{-i(\psi + 2\Delta_j)}\right), \quad (7)$$

$$E_d^{'}(t) = \frac{iE_0}{\sqrt{2}} e^{i(f_0\tau + i\Delta_j)} \left(1 + e^{-i(\psi + 2\Delta_j)}\right). \quad (8)$$

$$\langle I_c^{'}(t) \rangle = \frac{I_0}{2N+1} \sum_{j=-N}^{N} [1 - \cos(\psi + 2\Delta_j)], \quad (9)$$

$$\langle I_d^{'}(t) \rangle = \frac{I_0}{2N+1} \sum_{j=-N}^{N} [1 + \cos(\psi + 2\Delta_j)]. \quad (10)$$

$$\langle R_{cd}^{'}(\tau) \rangle = \frac{I_0^2}{2N+1} \sum_{j=-N}^{N} sin^2(\psi + 2\delta f_j \tau). \quad (11)$$

Likewise, $P_{cd}^{'}(0) = I_c^{'}(t) I_d^{'}(t) = I_0^2 sin^2(\psi + 2\delta f_j \tau) = P_{cd}(\tau)$. Compared with MZI 1, MZI 2 shows a perfect swapping in the first-order intensity correlations: $I_c(\tau) = I_d^{'}(\tau)$ and $I_d(\tau) = I_c^{'}(\tau)$. This relation is quite important for the definition of entanglement, i.e, product-basis superposition: $|X\rangle = \left(|+\Delta_j\rangle_u |-\Delta_j\rangle_l\right) \pm \left(|-\Delta_j\rangle_u |+\Delta_j\rangle_l\right)$. The subscripts indicate the upper and lower paths of MZIs.

From Eqs. (3), (4), (9), and (10), thus, the statistical averages of the grouped intensities from both MZIs result in $\langle \overline{I_c} \rangle = \frac{1}{2}\left(\langle I_c \rangle + \langle I_c^{'} \rangle\right) = I_0$ and $\langle \overline{I_d} \rangle = \frac{1}{2}\left(\langle I_d \rangle + \langle I_d^{'} \rangle\right) = I_0$, satisfying uniform intensities in the first-order intensity correlations [25,26]. Here, the phase basis of MZI 1 is $\varphi \in \{0, \pi\}$, resulting in photon bunching into either output port of MZI 1. By setting $\psi = \varphi \pm \pi$ in MZI 2, intensity swapping in the output ports is satisfied. However, the two-photon correlation has no difference: $\langle R_{cd}(\tau) \rangle = \langle R_{cd}^{'}(\tau) \rangle = \langle \overline{R_{cd}}(\tau) \rangle = \frac{1}{2}(\langle R_{cd}(\tau) \rangle + \langle R_{cd}^{'}(\tau) \rangle) = \frac{I_0^2}{2N+1} \sum_{j=-N}^{N} sin^2(\psi + 2\delta f_j \tau)$. As $|\tau|$ increases, the $\delta f_j \tau$-dependent ensemble effects become dominant, resulting in the heterodyning degradation. For $\tau \gg \Delta^{-1}$, $\langle R_{cd}(\tau) \rangle = I_0^2/2$ results in, which is the classical lower bound of $g^{(2)} = 0.5$. Thus, the coherence HOM model composed of two independent MZIs satisfies the general HOM effect based on entangled photon pairs. Here, the $\pi$ phase difference between two MZIs is equivalent to the $\pm \pi/2$ phase shift between entangled photons from SPDC.

It is clear that the classical intensity product between Eqs. (3) and (4) is not equal to Eq. (6), unless the gated measurements are conducted individually in a fast detection regime. This indicates a critical role of coincidence detection between frequency-correlated fields even in a classical regime [5]. If such a selective measurement is possible coherently, it would open the door to a new stage of quantum technology. In a coherence approach for independent and symmetric MZIs, quantum correlation $R_{cd}(\tau)$ is possible in a fast detection regime. Thus, Eqs. (3) and (4) might be directly used to calculate the classically defined intensity product $P_{cd}(\tau)$: $P_{cd}(\tau) = I_0^2 sin^2(\delta f_j \tau)$. In this case, the temporal resolution of a PD determines the resolving power of the delay time $\tau$ between $I_c$ and $I_d$. Such a coherence intensity product can be seen on a fast digital oscilloscope in a gated mode. Thus, coherent light-based intensity product $\langle P_{cd}(\tau) \rangle$ results in the same anti-correlation of the entangled photon-based HOM effects.

## 4. NUMERICAL CALCULATIONS

Figure 2 shows the numerical calculations of the analytical solutions of Eq. (6) for the HOM effect. The AOM-induced spectral bandwidth is set to $\Delta = 10^8$ $Hz$, whose corresponding decoherence time of the ensemble photons is $t_s = 10^{-8}$ s. Equations (1), (2), (7), and (8) are numerically calculated for both first- and second-order intensity correlations with the spectral range of $-2\Delta \leq \delta f_j \leq 2\Delta$. All individual photons have sinusoidal oscillations at different frequencies to satisfy the wave nature of a photon. As shown in Fig. 2, the mean intensities of the sum of individual output photons from BS-1 and BS-2 are shown to be uniform as expected, regardless of $\tau$ (see the blue lines). For $\varphi = 0$, the mean two-photon correlation $\langle R_{ab}(\tau) \rangle$ $(= \langle R'_{cd}(\tau) \rangle = \langle \overline{R_{cd}}(\tau) \rangle)$ satisfies the HOM dip with a $\tau$-dependent decay rate in both wings, as shown in the red curve in Fig. 2(a). This $\tau$-dependent decay is due to the ensemble decoherence given by $(\Delta)^{-1}$. The oscillation curve is the direct proof of the ensemble-averaged beating signal between output photons. If the resolving time of a photon detector is not good enough compared with the photon's bandwidth, such an oscillation cannot be seen [3]. For $\varphi = \frac{\pi}{2}$, however, a perfect super-Poisson in thermal photons is shown in Fig. 2(b). For $\varphi = \pi/4$, a uniform intensity is shown for the perfect individuality of photons in classical physics (see the red line in Fig. 2(c)). Thus, it is clear that the second-order intensity correlation of the HOM effect critically depends on $\varphi$, where this $\varphi$ determines the inherent phase shift $\eta$ between paired photons. This fixed phase relation between paired photons has never been discussed carefully in the quantum information community yet [1,3-12]. Thus, the quantum validity of the coherence model in Fig. 1 is successfully demonstrated for the HOM effect.

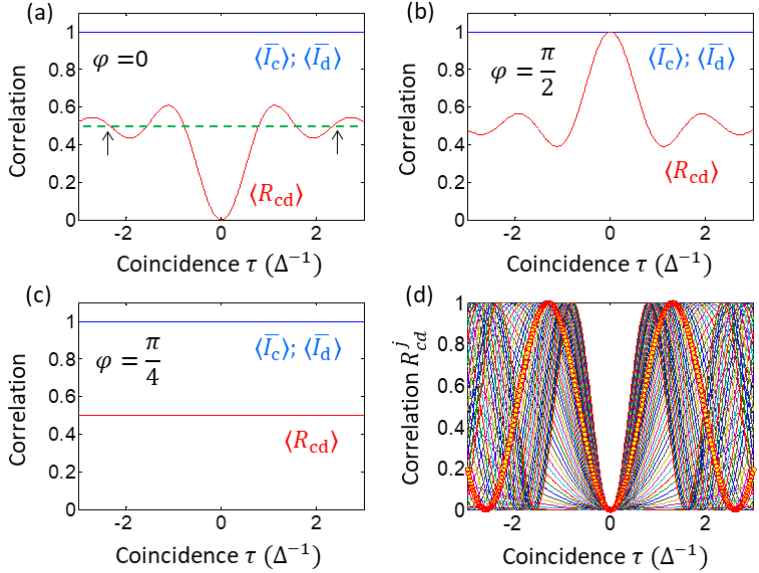

**Fig. 2. Numerical calculations for two-photon intensity correlation in Fig. 1**. (a)-(c) Overall two-photon intensity correlation for different $\eta$. (d) All $\delta f_j$-dependent individual two-photon intensity correlations for (a). The red circled curve is for a detuned pair indicated by the arrows in (a). For methods and parameters, see the text. $\varphi$ is the added phase shift to the upper path of MZI.

Figure 2(d) is for individual intensity products of Fig. 2(a). The red-circle curve is for a particularly detuned pair, whose detuning is indicated by arrows in Fig. 2(a). As shown, all individual intensity products are a square of the sine function, representing the origin of the HOM effect based on the coherence feature of individual photon pairs. In an ensemble average in Fig. 2(a), the two-photon intensity correlation gradually decreases even though individual correlations do not, as shown in Fig. 2(d). Regardless of detuning $\delta f_j$, the two-photon intensity product is always zero at $\tau = 0$ for a fixed relative phase $\eta = \frac{\pi}{2}$ ($\varphi = 0$). Thus, the HOM dip with no optical interference fringes is understood as an $\delta f_j \tau$-dependent ensemble decoherence of independently measured intensity products.

## 5. DISCUSSION

In both MZIs of Fig. 1, the AOMs are controlled by continuous (or pulsed) rf frequencies. Thus, the free-running time 't' is added to the delay time $\tau$. On the other hand, SPDC-generated entangled photons, i.e., signal and idler photons have no such free-running time. This inevitable free-running time 't' by AOM modulators is serious to a single-photon counting module due to its averaging of the beating signals in a time domain. As a result, no beating fringe is obtained for both first- and second-order intensity correlations in each MZI. To solve this matter, the single photon counting module is replaced by a fast digital oscilloscope to freeze the free-running time in a gated mode operation, Thus, freezing its measurement time can retrieve the role of delay time $\tau$. For example, the free-running beating signal on a fast digital oscilloscope can the gated for a narrow time slot, resulting in $\tau$-dependent intensity fringes from each MZI output port. Once the gated time window is fixed in the fringe period, both measured signals can be used for the second-order intensity correlations via an intensity product function. For this, fast PDs can replace the single-photon detectors for a pulse scheme to isolate each measurement event. The bandwidth of PDs must be much wider than the AOM bandwidth for the gating operation. For a pair of fixed $\pm \Delta_j$ of AOMs, near-perfect fringe visibility is obtained for $\langle \overline{R_{cd}}(\tau) \rangle$, $\langle I_c \rangle$ and $\langle I_d \rangle$. Due to the benefit of coherence optics, thus, a macroscopic quantum correlation can also be exploited. Experimental demonstrations of the macroscopic HOM effects are discussed elsewhere.

## 6. CONCLUSION

Using coherently manipulated photon pairs from an attenuated laser, the HOM effect was analyzed in a pair of symmetric MZIs for a coherence HOM model, and numerical demonstrations were demonstrated for the same quantum features observed with entangled photons. In the proposed coherence HOM model, a uniform value of the first-order intensity correlations resulted from the basis sum of both MZIs. The photon bunching phenomenon of the HOM effect was due to a fixed phase relation between paired photons on a BS. This understanding is rooted in the self-interference of a single photon. For the coherence HOM model, symmetrically modulated frequency-path correlations were adopted to mimic the SPDC process. With synchronized AOM manipulations, a HOM dip with ensemble-averaged beating fringes was analytically and numerically verified for the mean value of the second-order intensity correlations. Varying the fixed phase, however, the quantum feature of the HOM effect was transferred to classical features. Unlike the particle nature-based understanding, the origin of the HOM effect was found in conventional coherence optics via selective measurements in a time domain, where coincidence measurements required a wider bandwidth of a detector than the photon bandwidth. In that sense, a coherence approach for the HOM effect could be expanded into a single-shot measurement of the HOM effect if the photodetector's bandwidth surpasses the photon bandwidth in a macroscopic scheme. The novelty of the present paper was in the complete coherence understanding of the HOM effect in terms of random product bases of MZI. Like general coincidence detection-caused measurement modification [31,32], the coherence approach of the HOM effect could be implemented by classical means.

**METHODS**

For the symmetric $\pm \delta f_j$ in the Inset of Fig. 1, a pair of synchronized AOMs or electro-optic modulators can be used in a fast frequency sweeping mode compared to the photon detector's bandwidth. To satisfy the randomness between paired photons, a pair of symmetrically modulated MZIs is used. Due to the benefit of a narrow linewidth (~MHz) of a laser at $f_0$ in frequency, even a commercial bandwidth $\Delta$ (~100 MHz) of the AOMs does satisfy the HOM condition of the random frequency-path correlations, where the corresponding path-length difference is ~6 m. Considering a dead time of a single photon detector at ~20 ns, a GHz scan speed of a modulator is good enough for the gated detection scheme. Taking opposite AOM diffractions in each MZI automatically solves the spatial overlap issue on a BS in a frequency sweeping mode.

Regarding freezing the free running time 't' in measurements, a gated measurement technique is applied to the digital oscilloscope directly connected by single-photon detectors. The gated measurement can also be conducted in a pulsed scheme. Because there is no difference between a single photon and cw in an MZI, the rf pulses applied to AOMs result in predetermined gated signals to be measured. As the duration of rf pulses, the role of gated measurements fades out.

For Fig. 2, Eqs. (1) and (2) are used for $-10^8 \leq \delta f \leq 10^8$ (Hz) and $-3\times10^{-8} \leq \tau \leq 3\times10^{-8}$ (s). The increment of $\delta f$ ($\tau$) is $2*10^6$ ($2*10^{-10}$). For Figs. 2(a)-(c), each individual intensity pair is calculated independently. For simplicity, a Gaussian distribution is not applied to the photon bandwidth. As shown in Fig. 2(d), individual intensity products show a square of the sine function, whose center value at $\tau = 0$ depends on φ. Time-dependent decoherence is not applied for the individual product calculations due to the small range of $\tau$ compared with the given coherence time of individual photons.

**Data availability**

The data presented in the figures of this Article are available from the corresponding author upon reasonable request.

**Code availability**

All custom code used to support claims and analysis presented in this Article is available from the corresponding author upon reasonable request.

**Funding:** This research was supported by the MSIT (Ministry of Science and ICT), Korea, under the ITRC (Information Technology Research Center) support program (IITP-2023-2021-0-01810) supervised by the IITP (Institute for Information & Communications Technology Planning & Evaluation). BSH also acknowledges that this work was supported by GIST via GRI 2023.
**Competing interests**
The authors declare no competing interests.

**Reference**
1. Bouchard, F., Sit, A., Zhang, Y., Fickler, R., Miatto, F. M., Yao, Y., Sciarrino, F. & Karimi, E. Two-photon interference: the Hong-Ou-Mandel effect. *Rep. Prog. Phys.* (2021) 84:012402.
2. Nielsen, Michael, M. A. & Chuang, I. L. *Quantum Computation and Quantum Information*. (Cambridge University Press, NY, 2000).
3. Hong, C. K., Ou, Z. Y. & Mandel, L. Measurement of subpicosecond time intervals between two photons by interface. *Phys. Rev. Lett.* (1987) 59:2044-2046.
4. Kaltenbaek, R., Blauensteiner, B., Zukowski, M., Aspelmeyer, M. & Zeilinger, A. Experimental interference of independent photons. *Phys. Rev. Lett.* (2006) 96:240502.
5. Lettow, R. *et al*. Quantum interference of tunably indistinguishable photons from remote organic molecules. *Phys. Rev. Lett.* (2010) 104:123605.
6. Lopez-Mago, D. & Novotny, L. Coherence measurements with the two-photon Michelson interferometer. *Phys. Rev. A* (2012) 86: 023820.
7. Lang, C., Eichler, C., Steffen, L., Fink, J. M., Woolley, M. J., Blais, A. & Wallraff, A. Correlations, indistinguishability and entanglement in Hong-Ou-Mandel experiments at microwave frequencies. *Nature Phys.* (2013) 9:345-248.
8. Kobayashi, T. *et el*., Frequency-domain Hong-Ou-Mandel interference. *Nature Photon*. (2016) 10:441-444.
9. Deng, Y.-H. *et al*. Quantum interference between light sources separated by 150 million kilometers. *Phys. Rev. Lett.* (2019) 123:080401.
10. Edamatsu, K., Shimizu, R. & Itoh, T. Measurement of the photonic de Broglie wavelength of entangled photon pairs generated by spontaneous parametric down-conversion. *Phys. Rev. Lett.* (2002) 89:213601.
11. ThomasR, . J., Cheung, J. Y., Chunnilall, C. J. & Dunn, M. H. Measurement of photon indistinguishability to a quantifiable uncertainty using a Hong-Ou-Mandel interferometer. *Appl. Phys.* (2010) 49:2173-2182.
12. Poulios, K., Fry, D., Politi, A., Ismail, N., Worhoff, K., O'Brien, J. L. & Thompson, M. G. Two-photon quantum interference in integrated multi-mode interference devices. *Opt. Exp.* (2013) 21:23401-23409.

13. Gerry, C. C. & Knight, P. L. *Introductory to Quantum Optics.* (Cambridge Univ. Press, Cambridge, 2005), ch. 6.
14. Dirac, P. A. M. *The principles of Quantum mechanics* (4th ed., Oxford university press, London, 1958), ch. 1, p. 9.
15. Degiorgio, V. Phase shift between the transmitted and the reflected optical fields of a semireflecting lossless mirror is π/2. *Am. J. Phys.* (1980) 48:81–82.
16. Ham, B. S. The origin of anticorrelation for photon bunching on a beam splitter. *Sci. Rep.* (2020) 10:7309.
17. Ham, B. S. Coherently controlled quantum features in a coupled interferometric scheme. Sci. Rep. (2021) 11:11188.
18. Kim S. & Ham, B. S. Observations of the delayed-choice quantum eraser using coherent photons. Sci. Rep. 13, 9758 (2023).
19. Franson, J. D. Bell inequality for position and time. *Phys. Rev. Lett.* 62, 2205-2208 (1989).
20. Kwiat, P. G., Steinberg, A. M. & Chiao, R. Y. High-visibility interference in a Bell-inequality experiment for energy and time. *Phys. Rev. A* (1993) 47:R2472–R2475.
21. Ham, B. S. The origin of Franon-type nonlocal correlation. arXiv:2112.10148v4 (2023).
22. Einstein, A., Podolsky, B. & Rosen, N. Can quantum-mechanical description of physical reality be considered complete? *Phys. Rev*. **47**, 777-780 (1935).
23. Bell, J. S. On the Einstein Podolsky Rosen paradox. *Physics* **1**, 195 (1964).
24. Greenberger, D. M., Horne, M. A. & Zeilinger, A. Multiparticle interferometry and the superposition principle. *Phys. Today* **46** (8), 22-29 (1993).
25. Cruz-Ramirez, H., Ramirez-Alarcon, R., Corona, M., Garay-Palmett, K.& U'Ren, A. B., Spontaneous parametric processes in modern optics. *Opt. Photon. News* (2011) 22:36-41, and reference therein.
26. Zhang, C., Huang, Y.-F., Liu, B.-H., Li, C.-F. & Guo, G.-C. Spontaneous parametric down-conversion sources for multiphoton experiments. *Adv. Quantum Tech.* (2021) 4:2000132.
27. Boyd, R. W. *Nonlinear Optics* (Academic Press, San Diego, 2003), ch. 2.
28. Kim, S. & Ham, B. S. Revisiting self-interference in Young's double-slit experiments. Sci. Rep. **13**, 977 (2023).
29. Solano, E., Matos Filho, R. L. & Zagury, N. Deterministic Bell states and measurement of motional state of two trapped ions. *Phys. Rev. A* (1999) 59:R2539–R2543.
30. Grangier, P., Roger, G. and Aspect, A. Experimental evidence for a photon anticorrelation effect on a beam splitter: A new light on single-photon interferences. Europhys. Lett. **1**, 173-179 (1986).
31. Ham, B. S. The origin of Franson-type nonlocal correlation. arXiv:2112.10148v4 (2023).
32. Ham, B. S. Coherence interpretation of nonlocal quantum correlation in a delayed-choice quantum eraser. arXiv: 2206.05358v4 (2023).